\documentclass[a4paper,twocolumn,prl]{revtex4-2}
\usepackage[latin9]{inputenc}
\usepackage[active]{srcltx}
\usepackage{color}
\usepackage{amsmath}
\usepackage{amssymb}
\usepackage{graphicx}

\makeatletter

\usepackage{bm}
\usepackage{lipsum}

\makeatother

\begin{document}
\title{Instability of QBC systems to Topological Anderson Insulating phases}
\author{Nicolau Sobrosa$^{1}$, Miguel Gonçalves$^{2}$, Eduardo V. Castro$^{1,3}$}
\affiliation{$^{1}$Centro de Física das Universidades do Minho e Porto, Departamento
de Física e Astronomia, Faculdade de Ciências, Universidade do Porto,
4169-007 Porto, Portugal}
\affiliation{$^{2}$CeFEMA, Instituto Superior Técnico, Universidade de Lisboa,
Av. Rovisco Pais, 1049-001 Lisboa, Portugal}
\affiliation{$^{3}$Beijing Computational Science Research Center, Beijing 100084,
China}
\begin{abstract}
Here we study the instabilities of a quadratic band crossing system
to Chern insulating states and uncorrelated disorder. We determine
the phase diagram in the plane of topological mass versus disorder
strength, characterizing the system with respect to spectral, localization
and topological properties. In the clean limit, the system has two
gapped Chern insulating phases with Chern numbers $C=\pm2$, and a
trivial phase with $C=0$. For finite disorder, the quadratic band
crossing points are unstable to emergent gapless Chern insulating
phases with $C=\pm1$, not present in the clean limit. These phases
occupy a considerable region of the phase diagram for intermediate
disorder and show features of topological Anderson insulators: it
is possible to reach them through disorder-driven transitions from
trivial phases.
\end{abstract}
\maketitle

\section{Introduction}

Topological insulators are a remarkable state of electronic matter.
They show quantized responses that are proportional to topological
invariants and, as a consequence, typically very robust to perturbations
and system's details \citep{RevModPhys.82.3045,QZrmp11,bernevigBook,Chiu2016}.
Topological band insulators, as paradigmatic examples of systems with
non-trivial topology, have been extensively studied and are fairly
well understood \citep{bernevigBook}. Non-trivial topology, however,
also manifests in systems with broken translational invariance that
are not described by topological band theory. Among these systems,
disordered topological insulators are a popular sub-group \citep{Wu_2016}.

Topological phases are robust to disorder regarding that no symmetry
protecting the topological properties is broken \citep{AZ97,Xiao2010}.
In quantum Hall insulators, disorder even plays a fundamental role
for the observation of a quantized Hall conductance. More generally,
it is now well established that in the case of Chern insulators, where
time-reversal symmetry is broken, uncorrelated disorder localizes
every eigenstate except at specific energies \citep{KM93,OMN+03,nagaosaQSHloc07,Castro2015,Castro2016}.
The extended eigenstates at these energies carry finite Chern numbers
and are therefore responsible for a quantized Hall response as long
as the Fermi level lies between them (the localized states cannot
change this response). Topological phase transitions in disordered
Chern insulators occur when the extended states merge and annihilate
at the Fermi level (through the so-called ``levitation and annihilation''
mechanism), becoming localized \citep{nagaosaQSHloc07}.

The localization properties of non-interacting topological systems
can be understood within a low energy description in terms of random,
massive Dirac Hamiltonians \citep{morimoto-alattocs2015}. Generic
phase diagrams in the plane of Dirac mass versus disorder strength
have been obtained for all ten symmetry classes from the tenfold way
\citep{schnyder2008classification}. For class~A, to which Chern
insulators belong, the phase diagram consists of multiple localized
phases which can be distinguished by their Chern number and are separated
by phase boundaries at which the localization length diverges. However,
not all Chern insulators can be derived from massive Dirac Hamiltoninans
at low energies. A well known example are quadratic band crossing
(QBC) systems.

Systems with quadratic band crossings in two-dimensions (2D) are very
interesting because, contrary to conventional band degeneracy points,
they are associated with a finite Berry phase of $\pm2\pi$. Due to
the finite density of states at the QBC, these systems are unstable
to interactions \citep{SYF+09}: originating nematic phases with two
Dirac cones, each carrying half of the QBC's Berry phase; or gap openings
that may give rise to topological insulating phases precisely due
to the non-trivial Berry phase of the QBC \citep{UH11,MV14,Ray2018,Zeng2018}.
The fate of interaction induced topological insulating phases in the
presence of disorder has been examined in Refs.~\citep{Wang2017,PhysRevB.102.134204}
within the one-loop RG approach. The suppression of topological phases
under increasing disorder and possible transition to trivial phases
was predicted.

Quite surprisingly, there are systems which show topological phase
transitions from trivial to topological phases with increasing disorder.
This disorder-driven topological phase is now known as topological
Anderson insulator (TAI) \citep{Shen2009,Groth2009}. This phase has
been observed in many models, including paradigmatic models of topological
insulators in 2D such as the Kane-Mele model \citep{KM05,Orth2016}
and the disordered Haldane model \citep{Song2012,Garcia2015,Goncalves2018},
in the 1D SSH model \citep{Liu2022}, in quasiperiodic systems \citep{Tang2022,Goncalves2022},
and more recently in non-Hermitian models \citep{Zhang2020}. Whether
TAI phases may be realized for Chern insulators derived from QBC systems
is still an open question. Moreover, the fate of the QBC itself in
the presence of disorder has received little attention.

\begin{figure}[t]
\centering{}\includegraphics[width=0.95\columnwidth]{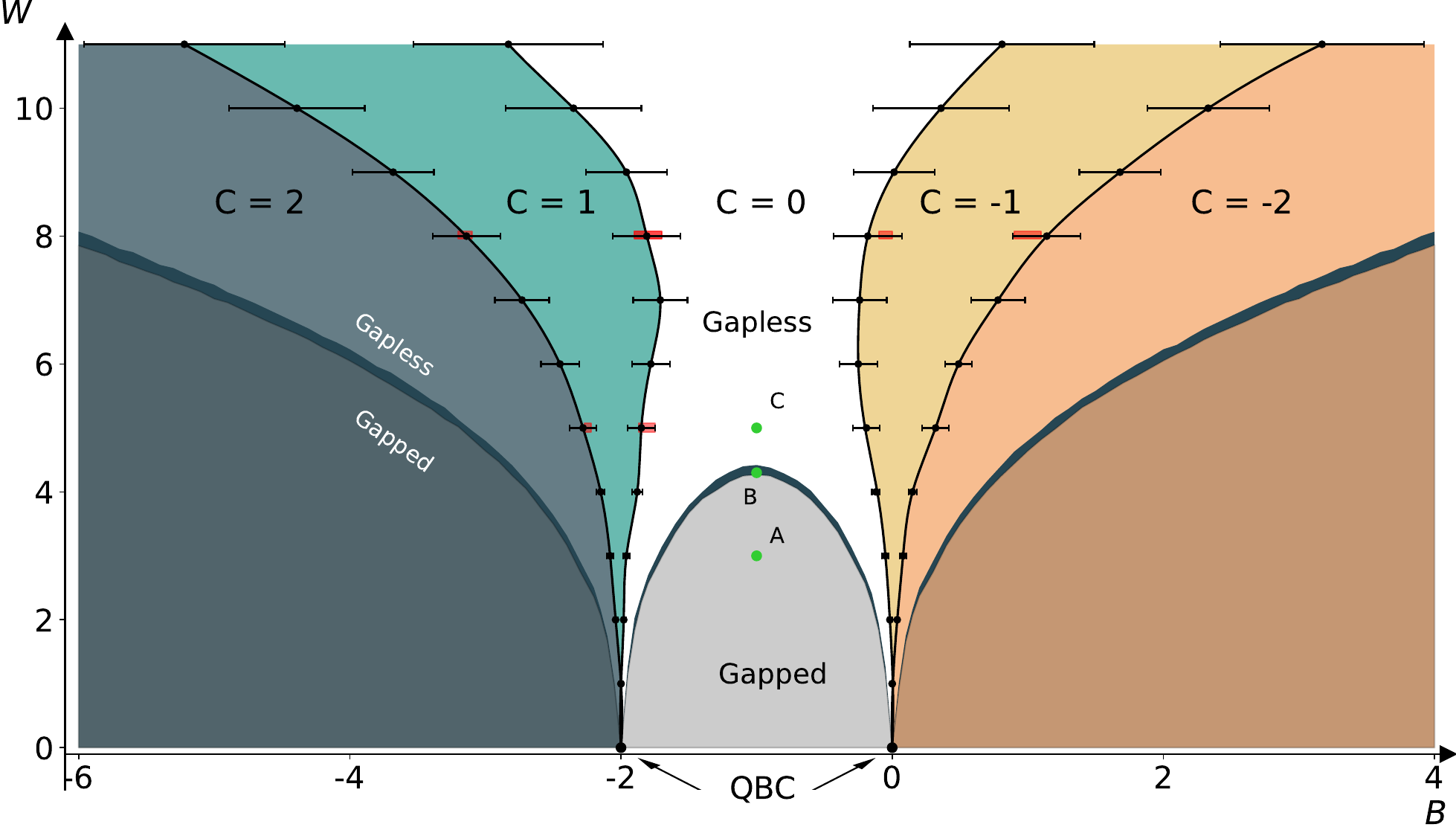}\caption{Phase diagram in the $B-W$ plane. The Chern number was calculated
for a system with $35\times35$ lattice points and was averaged over
$100$ disorder realizations. The gray areas, below the thicker lines,
correspond to the gapped regions obtained through calculations of
the DOS. The error bars for the Chern number were obtaining by fixing
the disorder intensity and varying $B$. The red boxes indicate the
results for the critical points calculated with the TMM. The size
of the bar is determined in such a way that outside the bar the system
is clearly localized ($\Lambda_{M}$ decreases with $M$, see Fig.~\ref{fig:Normalized-localization-length}
and main text).\label{fig:Phase-diagram}}
\end{figure}

In this work we study the interplay between the instability of a 2D
QBC system to a Chern insulating state and disorder of the Anderson
type. The full phase diagram in the plane of the gap opening coupling
parameter $B$ and disorder strength $W$ is shown in Fig.~\ref{fig:Phase-diagram}.
For null disorder, QBC points (QBCP) occur for $B=-2,0$, while the
system is a gapped trivial insulator or a Chern insulator with Chern
number $C=\pm2$ for any other $B$ value. These phases are also present
for finite disorder, but new gapless Chern insulating phases also
emerge. Besides gapless and gapped phases with $C=\pm2$, new gapless
topological phases with $C=\pm1$, not present in the clean limit,
arise. In fact, the most important result of our work is that the
QBCP are unstable to the formation of these phases, for any finite
disorder. Finally, the existence of TAI phenomena is also clear: it
is possible to undergo a transition between the trivial phase and
the disorder-induced topological phases with $C=\pm1$ by increasing
disorder.

The paper is organized as follows: In Sec.~\ref{sec:Model-and-methods},
we introduce the tight-binding model used to describe the electronic
properties of the disordered QBC system and the methods to analyze
its properties. The topological, spectral, and localization properties
are discussed in Sec.~\ref{sec:results}. A thorough discussion of
the obtained results is given in Sec.~\ref{sec:discuss}. In Sec.~\ref{sec:Conclusions}
the key results are summarized and some conclusions are drawn. We
also include four appendices: in Appendix~\ref{secap:Binary-disorder}
we provide the phase diagram for binary disorder; in Appendix~\ref{secap:Gapped-gapless-regions}
we discuss the criterion used to distinguish gapped and gapless regimes;
an example of a possible $\mathbf{k}$-dependent self-energy that
would lead to a $C=1$ phase in the self consistent Born approximation
is given in Appendix~\ref{secap:Self-Consistent-Born}; the robustness
of the obtained phase diagram when the QBC is split into two Dirac
cones is discussed in Appendix~\ref{secap:Evolution-of-Phase-with-bx}.

\section{Model and methods\label{sec:Model-and-methods}}

We study a QBC system realized on the square lattice with two orbitals
per site. The model considers first-neighbor hoppings between the
same orbitals and second-neighbor hoppings coupling different orbitals.
The Hamiltonian for the disorder-free model can be written in the
reciprocal space as

\begin{equation}
H_{0}=\sum_{\mathbf{k}}\boldsymbol{\Psi_{\mathbf{k}}^{\dagger}}\mathcal{H}_{\mathbf{k}}\boldsymbol{\Psi_{\mathbf{k}}}\,,\label{eq:H_zero}
\end{equation}

\noindent where $\boldsymbol{\Psi_{\mathbf{k}}^{\dagger}}=\left(c_{\mathbf{k}1}^{\dagger},c_{\mathbf{k}2}^{\dagger}\right)$
is the two component spinor in the space of the two orbitals, where
$c_{\mathbf{k}\alpha}$ creates an electron with Bloch momentum $\mathbf{k}$
in orbital $\alpha$, and

\begin{equation}
\mathcal{H}_{\mathbf{k}}=\mathbf{h}\cdot\boldsymbol{\sigma}\,,\label{eq:Hk}
\end{equation}

\noindent with $\boldsymbol{\sigma}$ the Pauli vector and $\mathbf{h}$
the vector given by:
\begin{align}
h_{x} & =2t_{x}\sin k_{x}\sin k_{y}\nonumber \\
h_{y} & =0\nonumber \\
h_{z} & =2t_{z}\left(\cos k_{x}-\cos k_{y}\right).\label{eq:h_real}
\end{align}
In the following, we set $t_{x}=t_{z}=t$ and $t$ to unity.

This model has two QBCPs: at $\Gamma=\left(0,0\right)$ and $\text{M}=\left(\pm\pi,\pm\pi\right)$.
By adding a finite $h_{y}$, it is possible to open a gap. For a constant
$h_{y}$, the system is a trivial insulator. Similarly to the Haldane
Model \citep{Haldane1988}, we may add a $k-$dependent component
which allows to tune independently the gap at each QBCP. In the following
we use a simple choice which depends on a single parameter $B$:

\begin{equation}
h_{y}=1+\frac{B+1}{2}\left(\cos k_{x}+\cos k_{y}\right).\label{eq:B_hy_alteratiojn}
\end{equation}
This type of $k-$dependence implies the breaking of time-reversal
symmetry since, as seen from Eq.~\eqref{eq:Hk}, we have $\mathcal{H}_{k}\neq\mathcal{H}_{-k}^{*}$.
As shown below, the system will become a Chern insulator for some
intervals of $B$ values.

In real-space, the constant term corresponds to an intracell complex
hopping between different orbitals and the term with $k-$modulation
to a nearest neighbor complex hopping between different orbitals.
As the modification in Eq.$\,$\eqref{eq:B_hy_alteratiojn} does not
change $h_{x}$ and $h_{z}$, a QBCP still exists when $h_{y}=0$
as seen before. For $B=0$, there is a QBCP at $\text{M}$ and for
$B=-2$, there is a QBCP at $\Gamma$.

Adding the disorder potential, the Hamiltonian reads:

\begin{equation}
H=H_{0}+\sum_{i}\sum_{\alpha=1,2}\xi_{i\alpha}\hat{c}_{i\alpha}^{\dagger}\hat{c}_{i\alpha}\:,\label{eq:complete_hamilto}
\end{equation}

\noindent where $\xi_{i\alpha}$ are site-dependent potentials that
follow the uniform distribution (Anderson disorder),

\begin{equation}
P_{W}\left(\xi_{i\alpha}\right)=\frac{1}{W}\Theta\left(\left|\xi_{i\alpha}\right|-\frac{W}{2}\right)\,,\label{eq:dis_distribution}
\end{equation}

\noindent and $W$ defines the disorder strength. In Appendix~\ref{secap:Binary-disorder}
we present the phase diagram obtained for the case of binary disorder.
The result is very similar to that of Fig.~\ref{fig:Phase-diagram}
obtained for Anderson disorder, and the conclusions are qualitatively
the same.

\noindent We carried out a complete study of the phase diagram of
the model in Eq.$\,$\eqref{eq:complete_hamilto}, characterizing
spectral, topological and localization properties. The density of
states (DOS) was calculated for finite systems containing more than
$10^{6}$ sites using the Kite~\citep{Joao2020} quantum transport
software that has a very efficient implementation of the kernel polynomial
method (KPM). The topological phase diagram was obtained by computing
the Chern number through the Coupling Matrix Method introduced in
Refs.~\citep{Fukui2005} and~\citep{Zhang2013}. We note that, even
though the Chern number computed for each disorder realization is
an integer within numerical accuracy, the averaged Chern number may
change continuously and present non-quantized values due to finite
size efffects. The localization properties were characterized through
the Transfer Matrix Method (TMM) \citep{MK81,MacKinnon1983,hoffmann2012computational},
that also allowed to cross-check the Chern number results. This method
considers a finite system with a large longitudinal length $L$ and
a transverse width $M$ which we varied in order to find the localization
length for a given $M$, $\lambda_{M}$. The localization properties
were studied through the normalized localization length, $\Lambda_{M}=\lambda_{M}/M$,
in the following way: if $\Lambda_{M}$ decreases with $M$ the states
are localized in the thermodynamic limit which corresponds to an insulating
behaviour; on the other hand, if $\Lambda_{M}$ increases with $M$
the states are extended; a constant $\Lambda_{M}$ is characteristic
of critical states that appear at transition points between different
phases. We choose $L$ so that $\lambda_{M}$ is calculated with an
error below $1\%$. We note that the behaviour of $\Lambda_{M}$ can
capture topological phase transitions at finite disorder. As mentioned
in the introduction, the spectrum of finite-disorder Chern insulators
consists of localized states, except at specific energies where critical
states live. Topological phase transitions occur when these states
cross (and merge at) the Fermi level. Therefore, for a disordered
Chern insulator, $\Lambda_{M}$ should always decrease with $M$ except
at the topological phase transitions, where it becomes $M$-independent.

\begin{figure}
\centering{}\includegraphics[width=0.95\columnwidth]{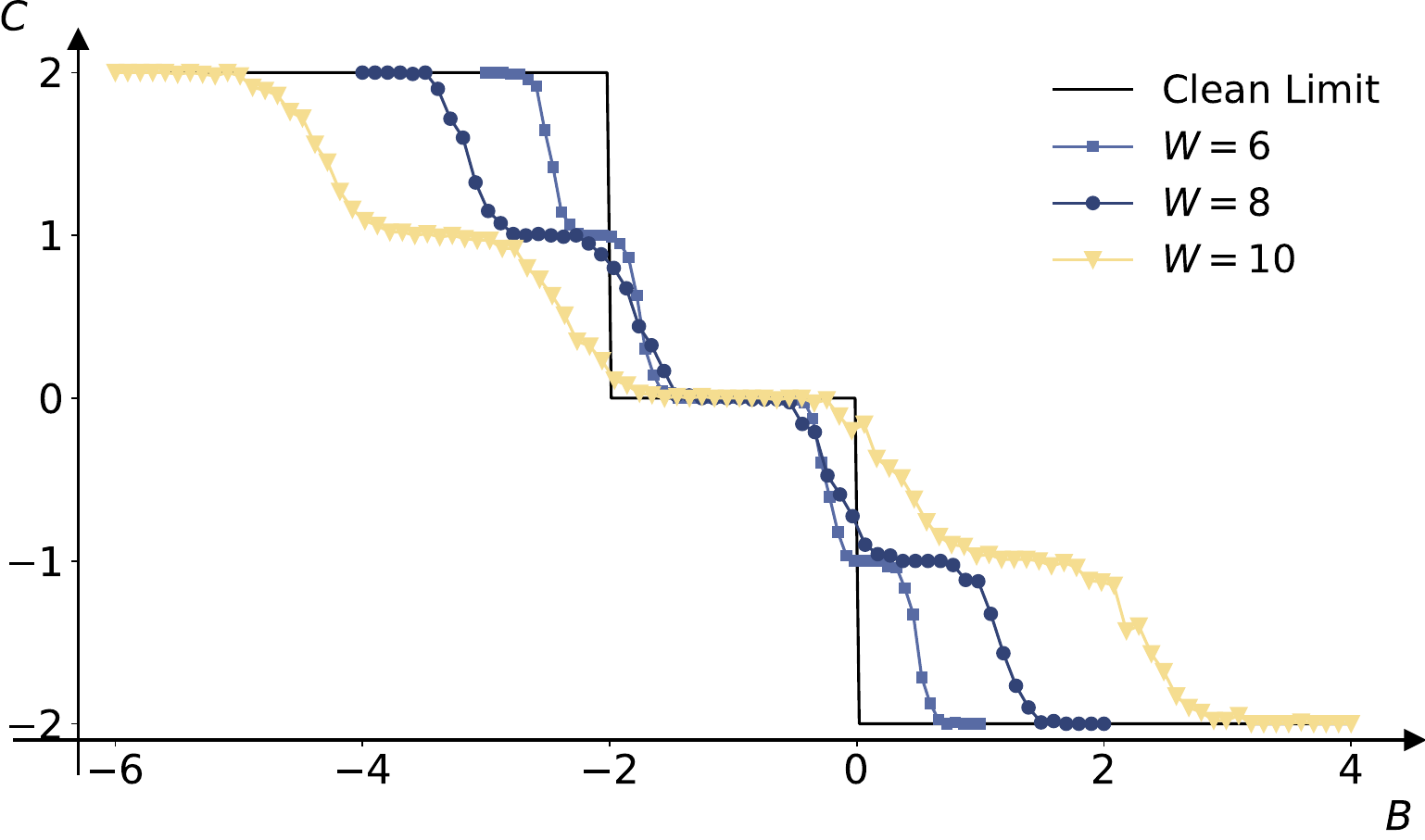}\caption{\label{fig:chern_curves_disorder}Chern number results obtained at
fixed $W$ and variable $B$. The results were averaged over 120 disorder
realizations. Each color depicts three examples of disorder intensities
used. The full black curve represents the topological phases of the
system calculated for null disorder where it is clear that no $C=\pm1$
phases exist.}
\end{figure}

\section{Results}

\label{sec:results}

\subsection{Topological properties}

In this section we present details on the topological phase diagram
of Fig.$\,$\ref{fig:Phase-diagram}. The different colors indicate
different Chern Numbers and the black, thin lines represent the topological
transitions. For null disorder the system undergoes a transition from
$C=\pm2$ to $C=0$ at the points where a QBCP appears, that is for
$B=-2$ and $B=0$. For finite disorder, the phases with $C=0,\pm2$
survive and new gapless phases with $C=\pm1$ appear. The latter phases
are TAI as it is possible to reach them by increasing disorder from
a topologically trivial phase, at fixed $B$. For large enough disorder,
all topological phases are suppressed, in agreement with one-loop
RG calculations \citep{Wang2017} for interacting QBC systems.

\begin{figure}
\begin{centering}
\includegraphics[width=1\columnwidth]{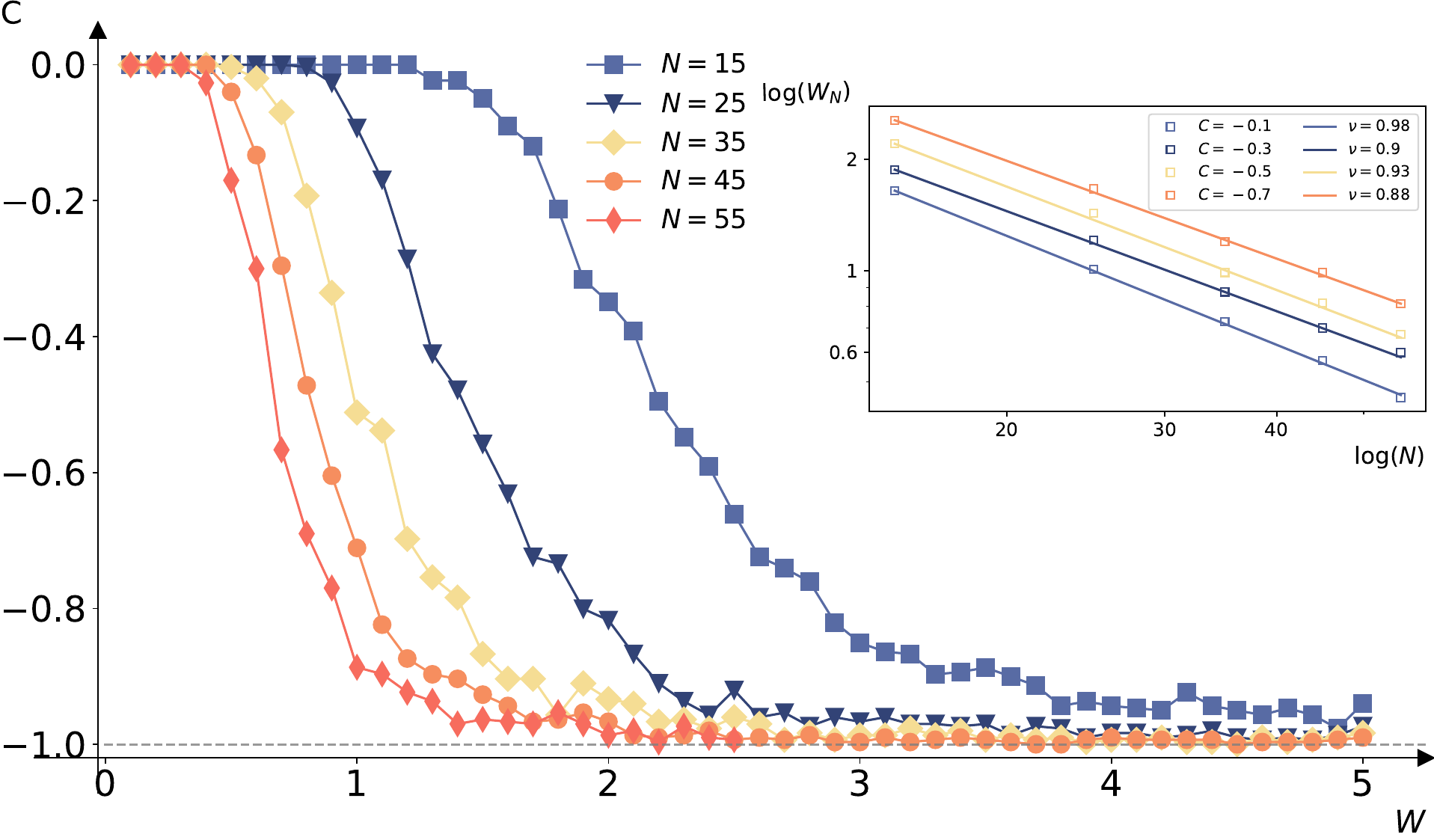}
\par\end{centering}
\caption{Chern number averaged over 300 disorder realizations as function of
disorder strength $W$ for $B=0$ (QBC at $W=0$) and variable system
sizes containing $N\times N$ sites. Inset: Scaling of some fixed
Chern numbers as function of the number of sites. $\nu$ indicates
the scaling as $N^{-\nu}.$\label{fig:QBC_instability}}
\end{figure}	

To obtain the transition lines, the disorder strength $W$ was fixed
at some value and the gap opening parameter $B$ was varied\textbf{
}continuously. For each disorder strength we performed an average
over $120$ different disorder configurations in a lattice with $35\times35$
sites. Some examples of the obtained curves are represented in Fig.~\ref{fig:chern_curves_disorder}
including the values for the clean limit. The continuous variation
of the averaged Chern number is expected to disappear in the thermodynamic
limit, where transitions between different Chern numbers should become
sharp. The error bars that appear in Fig.~\ref{fig:Phase-diagram}
are determined through the analysis of curves as the ones shown in
Fig.~\ref{fig:chern_curves_disorder}: the length of the bar indicates
the range over which the Chern number was more than $5$\% away from
its integer value.

From the phase diagram in Fig.$\,$\ref{fig:Phase-diagram}, it is
not clear if the finite-disorder Chern insulating phases with $C=\pm1$
exist for any infinitesimal disorder. In order to shed light on whether
this is the case, we fixed $B=0$ (QBC system in the clean limit)
and computed the Chern number as a function of $W$ for different
system sizes. The results are shown in Fig.$\,$\ref{fig:QBC_instability},
where it is clear that the $C=-1$ phase occurs for disorder strengths
that approach $W=0$ as the system size is increased. In the inset
of Fig.~\ref{fig:QBC_instability} we plot, for fixed $C$ values
in the apparent transition region, the corresponding disorder strength
$W_{N}$ for each size $N$. The fits clearly indicate that the transition
should occur discontinuously at $W=0$ in the thermodynamic limit.
Although we are limited numerically to reach larger system sizes,
these results support the conjecture that in the thermodynamic limit
the $B=0$ QBCP is unstable to the formation of the $C=-1$ phase
for any infinitesimal disorder. The same is expected for the QBCP
at $B=-2$, as suggested by the symmetry of the phase diagram around
$B=-1$.

\subsection{Gapped/Gapless regions}

\begin{figure}
\begin{centering}
\includegraphics[bb=0bp 0bp 530bp 265bp,width=1\columnwidth]{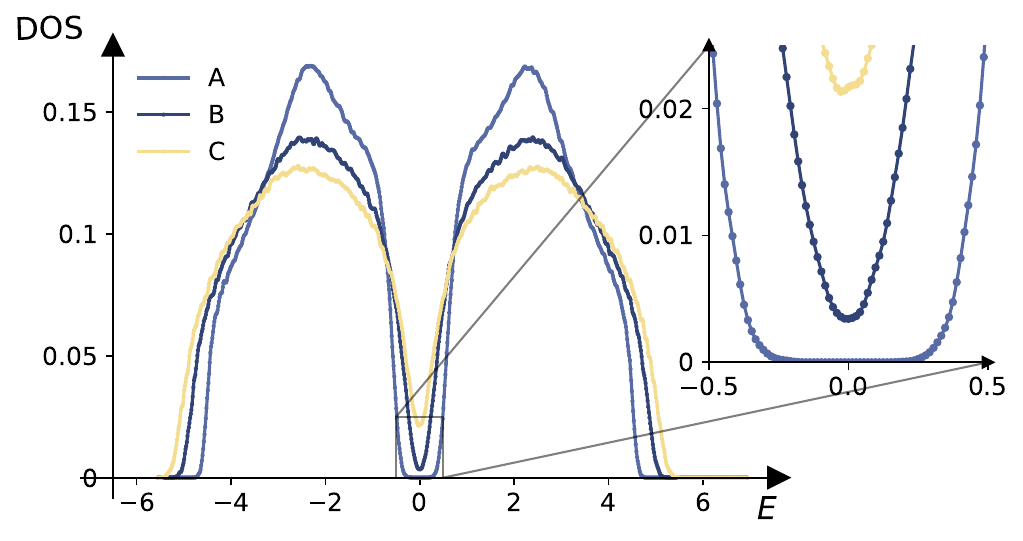}
\par\end{centering}
\caption{\label{fig:DOS_kite} Examples of DOS calculated for the points A,B,
and C, marked in Fig.~\ref{fig:Phase-diagram}. The inset shows a
zoom around the Fermi level which corresponds to $E=0$ at half-filling.
An average over $20$ disorder realizations was performed and a system
of $1024\times1024$ sites was considered.}
\end{figure}

In order to study the existence of a spectral gap at the Fermi level
($E=0$ at half-filling) we computed the DOS using KPM as implemented
in Kite \citep{Joao2020}. In Fig.~\ref{fig:Phase-diagram} we present
the gapped/gapless transitions with thicker lines, and shaded gapped
regions in grey. Examples of the DOS are shown in Fig.~\ref{fig:DOS_kite}
for systems with parameters corresponding to the points marked with
A, B, and C in the phase diagram of Fig.~\ref{fig:Phase-diagram}.
As seen in the inset, system~A is clearly gapped, with zero DOS at
and around $E=0$. System~C is gapless, as it presents a finite DOS
at and around $E=0$. Although system~B also presents a finite DOS,
its value at $E=0$ is small. In such cases, we used a criterion to
distinguish gapless from gapped systems, as explained next.

The system was considered gapped when the DOS at the Fermi Level was
below a certain threshold, $\rho_{\text{cut}}$, that was determined
by exact diagonalization of the Hamiltonian in Eq.~\eqref{eq:complete_hamilto}.
In Appendix~\ref{secap:Gapped-gapless-regions} we detail the analysis
we performed and show the results that support our choice. The error
in the transition from gapped to gapless regions corresponds to a
variation of $\pm25\%$ of $\rho_{\text{cut}}$. This error corresponds
to the thickness of the gapped-gapless transition lines in Fig.$\,$\ref{fig:Phase-diagram}.
The small error shows that variations in the criterion do not affect
significantly the results, especially in the regions of the topological
phase transitions.

\subsection{Localization properties}

\begin{figure}
\centering{}\includegraphics[width=0.48\textwidth]{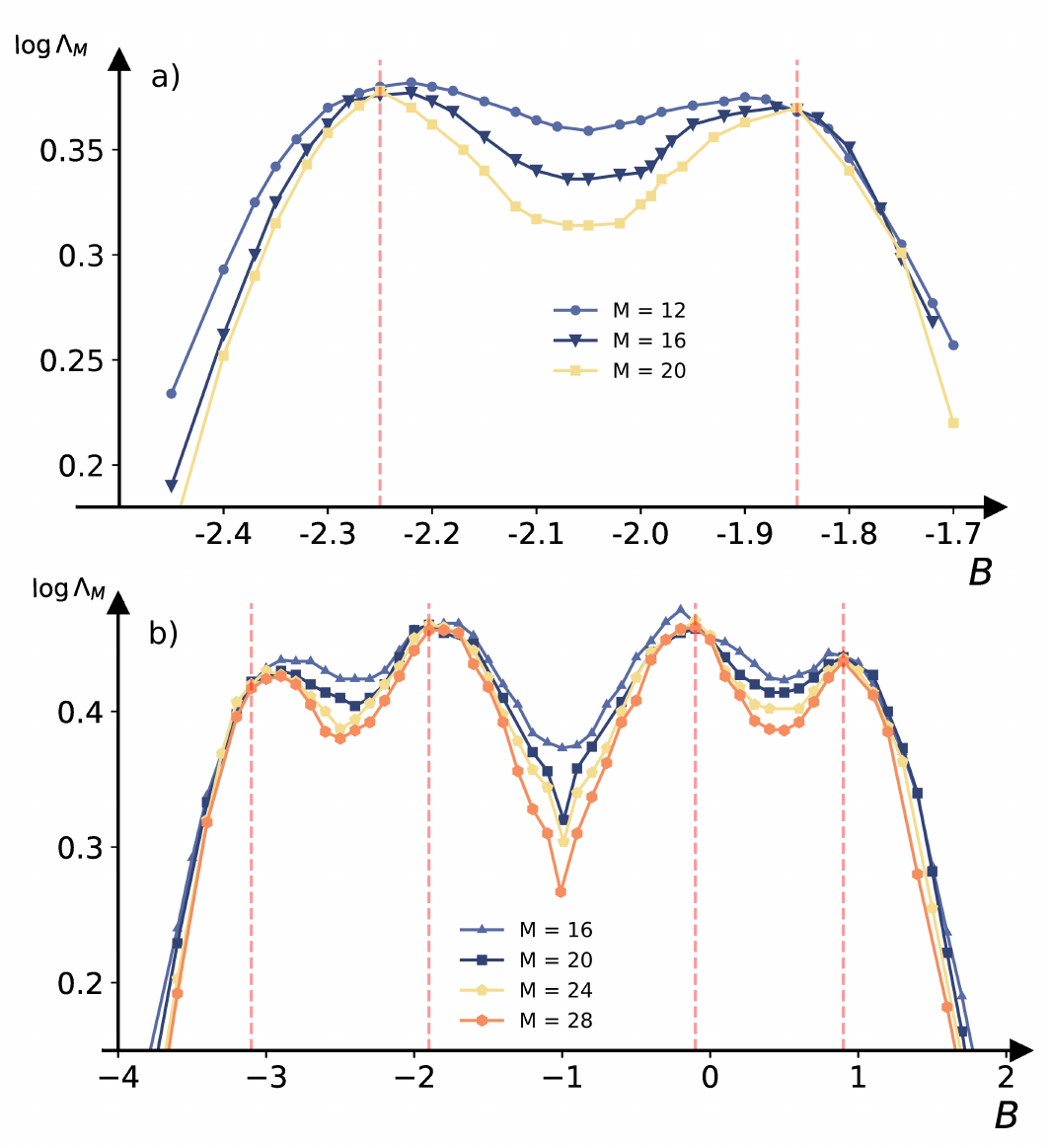}\caption{\label{fig:Normalized-localization-length}Normalized localization
length $\Lambda_{M}$ as a function of the gap opening parameter $B$,
for systems with varying width $M$ and disorder $W=5$ (a) and $W=8$
(b). The vertical dashed lines indicate the critical points.}
\end{figure}

We studied the normalized localization length $\Lambda_{M}$ obtained
through the TMM, at the Fermi level ($E=0$). We recall that in Chern
insulating systems, we expect $\Lambda_{M}$ to always decrease with
$M$, except at specific points. This means that, as expected, all
states at the Fermi level are localized except when a topological
phase transition occurs accompanied by the merging of critical states
carrying opposite Chern numbers. Example results are shown in Fig.$\,$\ref{fig:Normalized-localization-length}.
The results in Fig.$\,$\ref{fig:Normalized-localization-length}(a)
are for $W=5$ and a range of $B$ containing the $C=2$ to $C=1$
and $C=1$ to $C=0$ transitions. Fig.$\,$\ref{fig:Normalized-localization-length}(b)
corresponds to a cut at $W=8$, for a range of $B$ containing all
the topological phase transitions in the system. The phase transition
points were considered to be those with constant $\Lambda_{M}$. The
uncertainty in these points was considered to be the range of $B$
over which there is no clear decrease of $\Lambda_{M}$ with $M$.
This uncertainty is represented in Fig.$\,$\ref{fig:Phase-diagram}
by the shaded red bars, being perfectly bounded by the error bars
in the Chern number.

\section{Discussion}

\label{sec:discuss}

The phase diagram of the QBC system here unveiled shows many features
that are characteristic of well-known disordered topological insulators
\citep{OMN+03,nagaosaQSHloc07,Castro2015,Castro2016,Shen2009,Groth2009,Orth2016,Song2012,Garcia2015,Goncalves2018,PhysRevLett.125.133603,Stutzer2018,Song_2015,PhysRevB.93.214206,PhysRevB.100.054108,PhysRevB.102.205425,PhysRevB.103.115430},
such as the existence of robust finite-disorder gapped and gapless
topological insulating phases. However, there is a feature that distinguishes
it from the previously explored models: the existence of new disorder-driven
topological phases with $C=\pm1$, absent in the clean limit. 

Particularly interesting is the plateaus transitions $C=\pm2\rightarrow\pm1\rightarrow0$
at finite disorder strength as the topological gap parameter $B$
is changed shown in Fig.~\ref{fig:chern_curves_disorder}. This shows
that an equivalent plateaus sequence is possible with increasing disorder.
This possibility is conjectured to be ruled out in the quantum Hall
systems \citep{hatsugai1999sum} and other Chern insulators derived
from Dirac (linear band crossings) Hamiltonians \citep{Song2016}.
In those systems, starting with $|C|\geq2$, a plateaus transition
$\Delta C=\pm1$ is never observed with increasing disorder due to
ensemble averaging over disorder realizations.

Our results suggest that the existence of a QBCP is a key ingredient
for the formation of these phases. In particular, the finite-size
scaling analysis in Fig.$\,$\ref{fig:QBC_instability} is a strong
indication that they are instabilities of the QBCPs: in the thermodynamic
limit, any infinitesimal amount of disorder should drive the QBC system
to one of these phases. Furthermore, from the phase diagram in Fig.$\,$\ref{fig:Phase-diagram},
it is clear that for low disorder the new topological phases are located
around the clean-limit QBCP.

The new phases with $C=\pm1$ are also TAI since they can be reached
by increasing disorder from a trivial phase. TAI phenomena were by
now observed in a multitude of disordered topological insulators.
However, there is an important difference from the conventional TAI
in the present case: the $C=\pm1$ TAI phases have a Chern number
that does not exist in any zero-disorder topological phase of the
model. Therefore, these TAI phases do not evolve smoothly from the
clean-limit phases as disorder is increased, in contrast to conventional
TAIs.

For conventional TAIs, a self-consistent, low-order Born approximation
is usually enough to capture the topological phase transition \citep{Groth2009,Yang2021}.
In the present case, since the $C=\pm1$ phases are not present at
zero-disorder, a perturbative approach is not well justified. Moreover,
within the Born approximation a $k$-independent self-energy is obtained,
which translates into a renormalization of the parameters of the original
model. For the Hamiltonian in Eq.~\eqref{eq:Hk}, no $k$-independent
self-energy of the general form $\Sigma=\boldsymbol{\mathbf{\Sigma}}\cdot\boldsymbol{\sigma}$,
with $\boldsymbol{\Sigma}=(\Sigma_{x},\Sigma_{y},\Sigma_{z})$, is
able to induce a $C=\pm1$ phase. Nevertheless, a $k$-dependent self-energy
$\Sigma(\mathbf{k})=\boldsymbol{\mathbf{\Sigma}}(\mathbf{k})\cdot\boldsymbol{\sigma}$
may give rise to the $C=\pm1$ phase, as shown in the Appendix~\ref{secap:Self-Consistent-Born}
for a particular example with $\boldsymbol{\mathbf{\Sigma}}(\mathbf{k})=(\Sigma_{x}(\mathbf{k}),0,0)$.
In Appendix~\ref{secap:Self-Consistent-Born} we even show that the
clean limit model with the proposed $k$-dependent self-energy is
adiabatically connected to the disorder induced $C=\pm1$ TAI we have
found for our QBC system. However, for the uncorrelated disorder used
here, no $k$-dependent self energy is allowed which invalidates such
an approach.

A possible variation of the QBC system is to split the QBCP into two
Dirac cones with a suitable perturbation (see Appendix~\ref{secap:Evolution-of-Phase-with-bx}).
Even in this case, a finite disorder gives rise to the $C=\pm1$ phases
which could be an argument against the importance of the QBCP for
their existence. Nonetheless, the topological information carried
by the split Dirac cones and the QBCP is the same since no gap is
opened in the splitting process. The topological properties of the
new system can then be traced back to the possibility of creating
a QBCP without closing the gap.

Finally, the results obtained here are not restricted to disorder
of the Anderson type. We also obtained the phase diagram for binary
disorder, being qualitatively similar to the phase diagram for Anderson
disorder (see App.~\ref{secap:Binary-disorder}). This indicates
that our conclusion on the instability of QBCP to TAI phases is robust
to model details.

\section{Conclusions}

\label{sec:Conclusions}

We have studied a model of a QBC system under gap-opening and disorder-inducing
couplings. A complete spectral, topological and localization analysis
was carried out in order to obtain a detailed phase diagram. We not
only found that the topological phases existing in the clean limit
were robust to disorder but also that new gapless topological phases
were formed. Most importantly, we found a new instability of the QBCP:
a disorder-induced instability to gapless topological phases with
Chern numbers $C=\pm1$, that are absent in the clean limit. The possibility
of emulating quantum Hamiltonians using ultracold gases of atoms in
an optical lattices \citep{Lewenstein2007,Bloch2008} opens interesting
prospects to realize the observed phases experimentally. In particular,
the ability to realize disordered or quasiperiodic potentials into
the system to induce localization phenomena has been recently achieved
\citep{Schreiber2015,Choi2016}.

An interesting question for future work is whether instabilities of
the QBCPs to electron-electron interactions can give rise to topological
phases with similar properties to the gapless topological insulators
here uncovered. The full phase diagram capturing the interplay between
disorder and interactions would then be a natural follow-up. 
\begin{acknowledgments}
The authors NS and EVC acknowledge partial support from Fundação para
a Ciência e Tecnologia (FCT-Portugal) through Grant No. UIDB/04650/2020.
MG acknowledges partial support from Fundação para a Ciência e Tecnologia
(FCT-Portugal) through Grant No. UID/CTM/04540/2019. MG acknowledges
further support from FCT-Portugal through the Grant SFRH/BD/145152/2019.
\end{acknowledgments}

\appendix

\section{Qualitative Phase Diagram for Binary Disorder}

\label{secap:Binary-disorder}

In this Appendix, we present the phase diagram for binary disorder.
In this case, the on-site potentials $\epsilon_{i}$ are randomly
generated according to the following distribution:

\[
P_{V}(\varepsilon_{i})=\frac{1}{2}\left(\delta\left(\varepsilon_{i}\right)+\delta\left(V-\varepsilon_{i}\right)\right).
\]

\begin{figure}
\centering{}\includegraphics[width=0.48\textwidth]{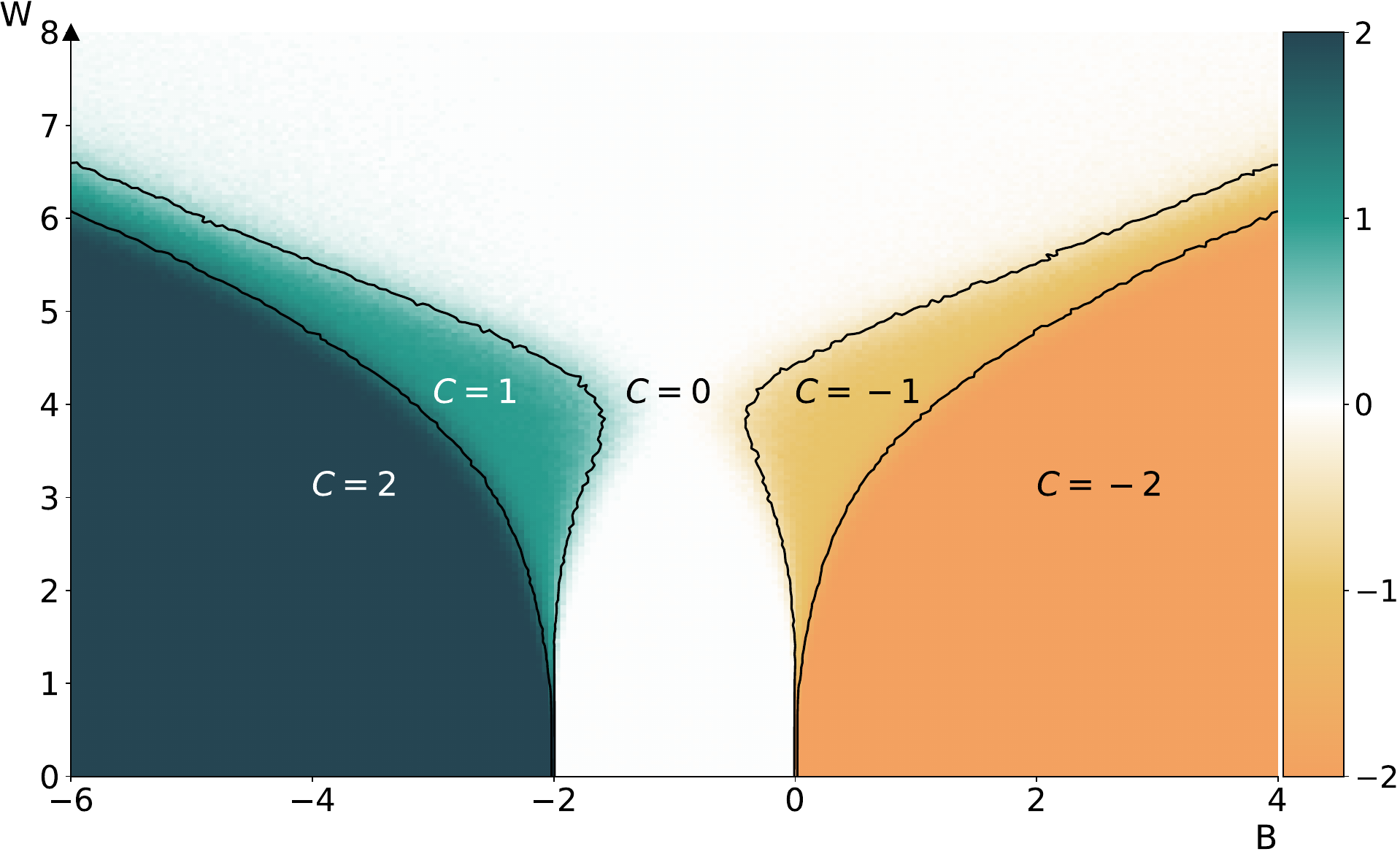}\caption{\label{fig:bin_phase}Phase diagram with Binary disorder. All points
were averaged over 200 disorder configurations and systems of size
$15\times15$ were used.}
\end{figure}

The Chern number results are shown in Fig.~\ref{fig:bin_phase}.
Qualitatively, the phase diagram is very similar to the one obtained
for Anderson disorder in Fig.$\,$\ref{fig:Phase-diagram}. In particular,
the disorder-induced $C=\pm1$ topological phases are still present
and the reentrant TAI behaviour is also observed. However, a smaller
degree of disorder is needed to destroy all the non-trivial phases.
Besides this, for large $B$ (absolute value), the $C=\pm1$ phases
are much narrower with binary disorder which means that they are more
robust to Anderson disorder.

\section{Gapped/gapless regions}

\label{secap:Gapped-gapless-regions}

In this Appendix, we present details on the criterion used in the
main text to distinguish gapped and gapless regimes.

\begin{figure}
\centering{}\includegraphics[width=0.48\textwidth]{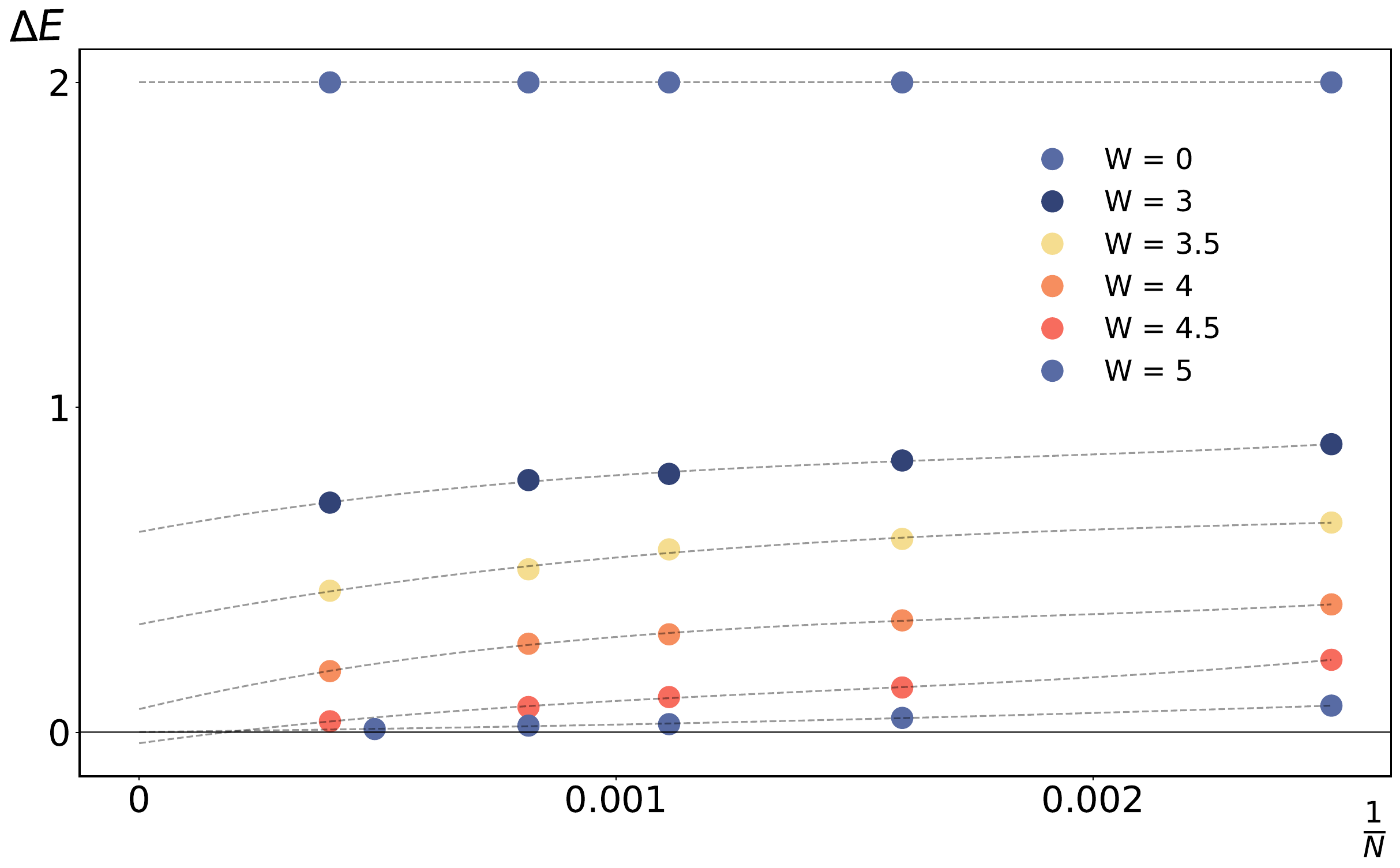}\caption{\label{fig:Energy-difference}Energy difference $\Delta E$ between
the two states closest to the Fermi level, directly above and below,
computed using exact diagonalization for systems with different disorder
strength $W$ and varying number of sites $N$.}
\end{figure}

Due to the finite resolution of KPM, it is typically challenging to
find the transition point between gapped and gapless regimes. In particular,
the DOS obtained through KPM may have a finite spectral weight at
energies within gaps if the system size and the number of polynomials
are not large enough. We therefore use finite-size scaling results
from exact diagonalization to find a suitable $\rho_{{\rm cut}}$
below which the KPM DOS should be considered null. For the system
to be gapless, the energy difference $\Delta E$ between the two states
closest to the Fermi level, immediately above and below, must converge
to zero in the thermodynamic limit, $\Delta E\rightarrow0$. To define
$\rho_{{\rm cut}}$, we set the parameter $B=-1$. Results for $\Delta E$
as a function of inverse system size $1/N$ are shown in Fig.~\ref{fig:Energy-difference}.
After fitting the finite-size scaling results to a cubic function
and extrapolating to $N\rightarrow\infty$, we observe that for a
disorder around $W\approx4.2$ the system must be gapless. It was,
then, just a matter of computing the DOS with Kite \citep{Joao2020}
for $B=-1$ with that critical disorder and then fix the $\rho(E=0)$
obtained with Kite to be $\rho_{cut}$. 

\section{Self Consistent Born Approximation}

\label{secap:Self-Consistent-Born}

\begin{figure}
\begin{centering}
\includegraphics[width=1\columnwidth]{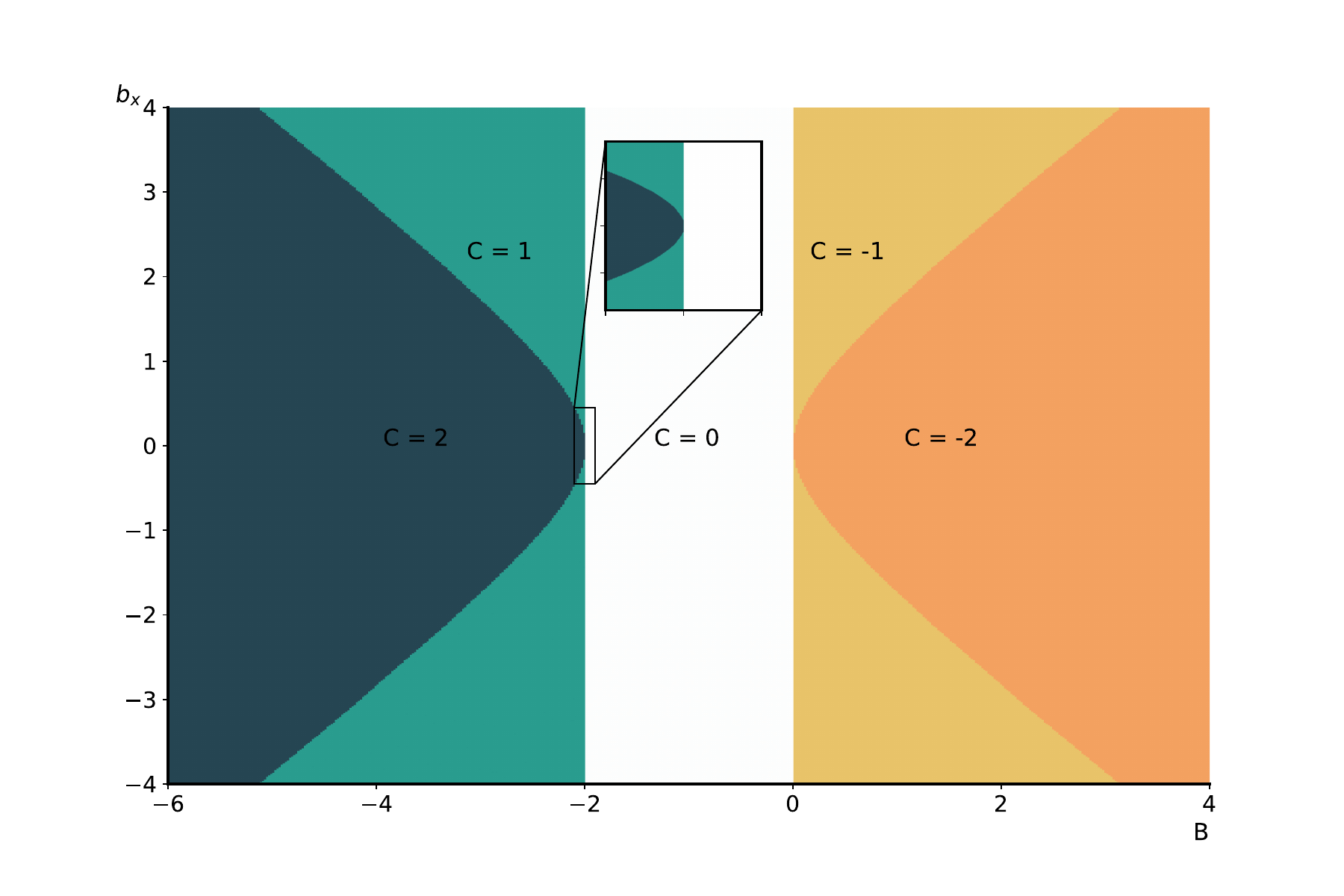}
\par\end{centering}
\caption{\label{fig:bxBchern}Example of a perturbation that leads to a C=1
phase without disorder.}
\end{figure}

Here we provide an example of a possible $\mathbf{k}$-dependent self-energy
term that leads to a $C=1$ phase:

\begin{equation}
\Sigma_{x}(\mathbf{k})=d_{x}\sin\left(k_{x}+k_{y}\right).\label{eq:selfEner}
\end{equation}
Results for the topological phases as a function of $B$ and $d_{x}$are
found in Fig.~\ref{fig:bxBchern}.

\begin{figure}
\begin{centering}
\includegraphics[width=1\columnwidth]{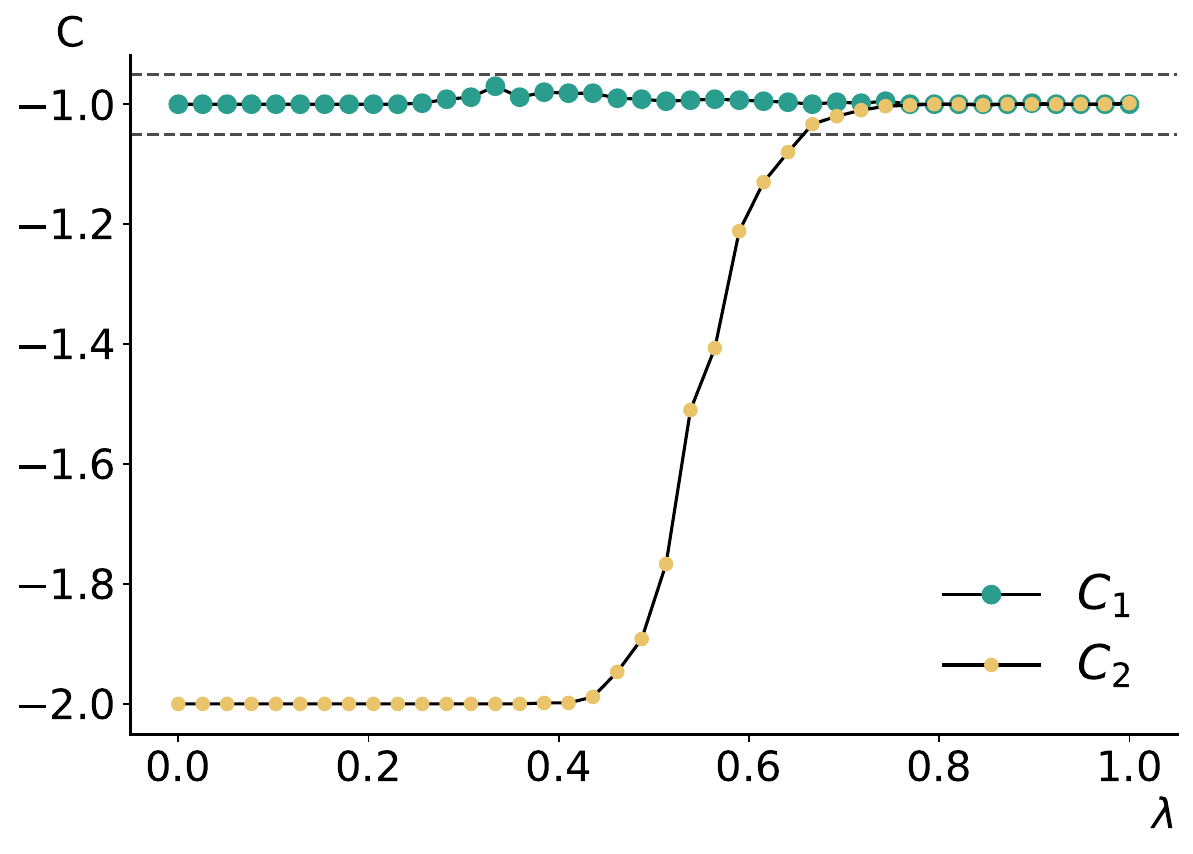}
\par\end{centering}
\caption{\label{fig:paths} Chern number computed along the paths $\mathcal{P}_{1}$
and $\mathcal{P}_{2}$ defined by Eqs.~\ref{eq:paths1} and~\eqref{eq:paths2},
respectively. All points were averaged over $600$ disorder realization
for system of size $45\times45$.}
\end{figure}

To see that these phases are adiabatically connected to the $C=\pm1$
disorder-induced phases we compute the Chern number in a parameter
path $\mathcal{P}_{1}$ given by the pair $(d_{x},W)$:

\begin{equation}
\mathcal{P}_{1}:\left(d_{x},W\right)=\begin{cases}
\left(0.7,12\lambda\right) & ,0<\lambda\leq\frac{1}{2}\\
\left(1.4(1-\lambda),6\right) & ,\frac{1}{2}<\lambda\leq1
\end{cases},\label{eq:paths1}
\end{equation}
where we have fixed $B$ to $B=0.1$. For $\lambda=0$, the model
has the self-energy given in Eq.~\eqref{eq:selfEner} and is in the
$C=-1$ phase of the phase diagram shown in Fig.~\ref{fig:bxBchern}.
When $\lambda=1$, we have $d_{x}=0$ which is a point of the phase
diagram in Fig.~\ref{fig:Phase-diagram} with $C=-1$. It can be
seen in Fig.~\ref{fig:paths} that the system has a $C=-1$ for all
points along the $\mathcal{P}_{1}$ path. The $\mathcal{P}_{2}$ path,
which is defined as
\begin{equation}
\mathcal{P}_{2}:\left(d_{x},W\right)=\left(0,6\lambda\right),\,\,\,\,0<\lambda\leq1,\label{eq:paths2}
\end{equation}
shows that the $C=-2$ phase may then be reached by decreasing disorder
$(\lambda=1\rightarrow0)$, in total agreement with the phase diagram
of Fig.~\ref{fig:Phase-diagram}.

\section{Evolution of Phase Diagram with $B_{x}$ constant perturbation}

\label{secap:Evolution-of-Phase-with-bx}

\begin{figure*}
\begin{centering}
\includegraphics[width=0.97\textwidth]{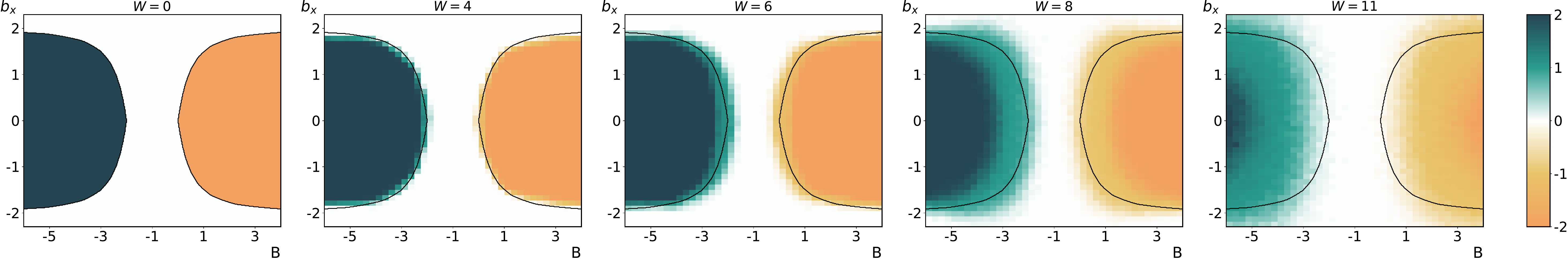}
\par\end{centering}
\caption{Evolution of phase diagram in the $B-b_{x}$ plane. The black lines
correspond to the phase transition for null disorder corresponding
to the points where there are Dirac Cones or QBCPs in the spectrum.
All points were averaged with 100 disorder configurations for systems
of size $15\times15$. \label{fig:Evolution-of-phase}}
\end{figure*}

To study the robustness of the QBCP we introduce the term $\boldsymbol{b}\cdot\boldsymbol{\sigma}$
in the Hamiltonian written in the reciprocal space where $\boldsymbol{b}$
can be a constant vector with three components, $\boldsymbol{b}=$$\left(b_{x},b_{y},b_{z}\right)$.
The $b_{x}$ perturbation lifts the degeneracy of the QBCP, splitting
it into two Dirac Cones. In fig. \ref{fig:Evolution-of-phase}, we
show the value of the Chern number in the plane \textbf{$B$ }vs\textbf{
$b_{x}$ }with increasing disorder. For a small $b_{x}$ the phase
diagram of the fig. 2 is practically unchanged. It is also clear from
the figure that the $C=\pm1$ phases exist even when $b_{x}\neq0$.

\bibliographystyle{apsrev}
\bibliography{mybib}

\end{document}